\documentstyle [12pt]{article}
\begin{document}
\baselineskip 20.0 pt
\par
\mbox{}
\vskip -1.25in\par
\mbox{}
 \begin{flushright}
\makebox[1.5in][1]{UU-HEP-92/7}\\
\makebox[1.5in][1]{August 1992}
 \end{flushright}
 \vskip 0.25in

\begin{center}
{\bf On the KP Hierarchy, $\hat{W}_{\infty}$ Algebra,
and Conformal SL(2,R)/U(1) Model}\\
{\bf I. ~ The Classical Case}\\
\vspace{40 pt}
{Feng Yu and Yong-Shi Wu}\\
\vspace{20 pt}
{{\it Department of Physics, University of Utah}}\\
{{\it Salt Lake City, Utah 84112, U.S.A.}}\\
\vspace{40 pt}
{\large ABSTRACT}\\
\end{center}
\vspace{10 pt}

In this paper we study the inter-relationship between the
integrable KP hierarchy, nonlinear $\hat{W}_{\infty}$ algebra and
conformal noncompact $SL(2,R)/U(1)$ coset model at the classical
level. We first derive explicitly the Possion brackets of the
second Hamiltonian structure of the KP hierarchy, then use it to define the
$\hat{W}_{1+\infty}$ algebra and its reduction $\hat{W}_{\infty}$.
Then we show that the latter is realized in
the $SL(2,R)/U(1)$ coset model as a hidden current algebra, through
a free field realization of
$\hat{W}_{\infty}$, in closed form for all higher-spin currents,
in terms of two bosons. An immediate consequence is the existence of
an infinite number of KP flows in the coset model,
which preserve the $\hat{W}_{\infty}$ current algebra.

\newpage
\section{Introduction}
\setcounter{equation}{0}
\vspace{5 pt}

This is the first of a two-paper series that give a unified
description of our recent results [1-4] on the close relationship
between the KP hierarchy and the $\hat{W}_{\infty}$ algebra, and
their field theoretical realization
in the 2d conformal non-compact $SL(2,R)/U(1)$ coset model. We will
provide detailed proofs for theorems we have stated in our previous
papers [1-4] and add some new results. In this paper
we concentrate on the classical aspects of the problem. The
generalization of the results to the quantum case and, in particular,
the problem of quantizing (or quantum deforming) the KP hierarchy are
left to the subsequent paper [5].

One of the most exciting recent developments in mathematical physics
is the revelation of the close relationship between nonlinear integrable
differential systems and 2d exactly solvable field theories, including
2d quantum gravity and string theory.  The essential links between them
turns out to reside in the common symmetry structure they share, which
usually is described by infinite dimensional algebras and ensures
the integrability of the differential systems, on one hand, and
the exact solvability of the field theories, on the other.
The first known example is the Virasoro algebra, the Lie algebra
of the infinite dimensional conformal symmetry in two dimensions.
On one hand, this algebra is known [6] to be isomorphic to the
second Hamiltonian structure of the KdV (Korteweg-de Vries)
equation or the KdV hierarchy. On the other hand, this algebra has
been shown [7] to be the symmetry algebra, or more precisely
the chiral algebra, of the so-called minimal conformal models in two
dimensions, in the sense that the fields in these models form a finite
number of irreducible representations (Verma modules) of this algebra.
The Virasoro algebra is realized in these models as the algebra of a spin-2
current, namely the energy-momentum tensor. Similar situations occur for
Zamolochikov's extended $W_N$ algebras ($N\geq 3$) [8]. Each of them is
an extension of the Virasoro algebra, which is generated by currents of
spin $2\leq s\leq N$; they are not Lie algebras, since there are
nonlinear terms in the commutators of generators, but all of them
contain the Virasoro algebra as a subalgebra. Again, it turns out that
the extended conformal $W_N$ algebra is isomorphic to the second
Hamiltonian structure of the $N$-th generalized KdV hierarchy [9], while
it is shown [10] to be the chiral algebra of certain conformal models
called the $W_N$ minimal models, in the sense that
fields in these models form a finite number of irreducible
representations of $W_N$.

Since $W_N$ exists for any integer $N\geq 2$, Bakas [11] first raised the
interesting question: Can we make sense of a large $N$ limit of the
$W_N$ algebras? By now we know the answer is yes but not unique,
depending on the way of defining the limit. All these ``large $N$ limits''
are now called the $W$ algebras of the infinite type or simply
$W$-infinity algebras. Two years ago, when we entered the field,
the list of known $W$-infinity algebras included:
1) a linear Lie algebra known as $w_{\infty}$
algebra obtained by Bakas [11];
2) a linear deformation of Bakas's $w_{\infty}$, known as $W_{\infty}$
algebra, proposed by Pope, Romans and Shen [12];
3) a further linear extension of $W_{\infty}$ by incorporating
an additional spin-1 current, again
by Pope, Romans and Shen [13] who named it as $W_{1+\infty}$.
All of them are (linear) Lie algebras. Though some simple free
field representations of these algebras were known [14,13], but no
connection to either integrable differential hierarchy or
2d conformal models had been made at that time.

It was very natural to us to try to generalize the above mentioned
profound relationship between (infinite dimensional)
extended conformal algebras, infinite integrable differential
hierarchies and 2d conformal field theories from $W_N$ with finite $N$
to $W$-infinity algebras. We started with studying the problem of
whether there is a ``large $N$ limit'' of $W_N$ which is related to
the KP (Kadomtsev-Petviashvili) hierarchy [15] in way as $W_N$
related to the generalized KdV hierarchy. The reason we choose the
KP hierarchy is twofold: First it is known [15] to contain all
$N$-th generalized KdV hierarchies by reduction; in some sense one may
view the KP hierarchy as a large $N$ limit of the generalized KdV
hierarchies. Secondly, the latter appear in recent matrix model [16] or
topological [17] approach
to 2d quantum gravity and string theory in two remarkable ways:
On one hand, the $N$-th KdV hierarchies can be used to formulate
the so-called string equations that determine all physical correlation
functions of a certain $c<1$, $D<2$ noncritical string [18].
On the other hand, the string partition function obeys a set of
constraints satisfying the $W_{N}$ algebra [19]. In either way, one needs
to consider an $N\rightarrow \infty$ limit to approach the famous
$c=1, D=2$ barrier. Therefore it is generally
believed that the KP hierarchy would play an significant role in
the study of the $c=1, D=2$ string theory.

The results we have reported in our previous papers [1-4] can be
summarized as follows:
1) The centerless $W_{1+\infty}$ can be identified
[1] (also see [20]) as the first Hamiltonian structure of
the KP hierarchy proposed by Watanabe [21]; an obvious reduction of the
latter gives rise to $W_{\infty}$ [1]. 2) It is proved [2] that
there exists a non-linear deformation of the $W_{\infty}$
algebra, which we call $\hat{W}_{\infty}$ and is believed to
contain all $W_N$ by reduction.
3) This $\hat{W}_{\infty}$ algebra is identified [2] as the second
KP Hamiltonian structure proposed by Dickey [22] (see also [23]).
4) We have found [3], in closed form for all higher spin currents,
a two boson realization of the nonlinear $\hat{W}_{\infty}$ algera.
5) Using it, we constructed a field theoretical realization of
$\hat{W}_{\infty}$ as a hidden current algebra in the 2d
$SL(2,R)/U(1)$ coset model.
6) In view of the well-known infinite set of KP Hamiltonians, considered
as $\hat{W}_{\infty}$ charges that are in involution, the result 5)
immediately implies the existence of infinitely many KP flows in the
coset model, all of which preserve the $\hat{W}_{\infty}$ current algerbra.
In the following sections we will give a unified description
and present the detailed
proofs for the classical theory. The treatemnt of the quantum theory is
left to the subsequent paper [5].

\vspace{30 pt}
\section{Bi-Hamiltonian Structure of the KP Hierarchy and
$\hat{W}_{1+\infty}$ and $\hat{W}_{\infty}$}
\setcounter{equation}{0}
\vspace{5 pt}

The integrable KP equation, originally proposed by Kadomtsev
and Petviashvili [24], was generalized by Sato [15] to an
integrable, infinite system, the so-called KP
hierarchy. This set of nonlinear partial-differential
equations can be nicely written in the form of a Lax pair,
if one introduces a first-order pseudo-differential operator
\begin{eqnarray}
L = D + \sum^{\infty}_{r=-1}u_{r}D^{-r-1}.
\end{eqnarray}
Here the coefficients $u_{r}$ are functions
of $z$ and various time variables
$t_{m} (m=1,2,3, \ldots)$, and $D\equiv \partial/\partial z$ obeys the
Leibniz rule $Df=fD+\partial_{z}f$ when acting on a regular function
$f$. Recall that an arbitrary pseudo-differential operator $O$
of order $N$ has the formal Laurent
expression
\begin{eqnarray}
O = \sum^{N}_{s=-\infty}o_{s}D^{s}.
\end{eqnarray}
The multiplication of two such operators is determined by the
generalized Leibniz rule. $D^{-1}$ is defined so that it
satisfies $D D^{-1} = D^{-1} D = 1$. It follows that, for positive $r$,
\begin{eqnarray}
D^{-r}u = \sum^{\infty}_{l=0}\left( \begin{array}{c}
                                      -r \\ l
                                    \end{array}  \right)
u^{(l)}D^{-r-l}
\end{eqnarray}
where $u^{(l)}\equiv \partial_{z}^{l}u$.
The KP hierarchy, written in terms of the operator (2.1), is a system
of infinitely many evolution equations for the functions $u_{r}
(r = 0,1,2,\cdots)$
\begin{eqnarray}
\frac{\partial L}{\partial t_{m}} = {[(L^{m})_{+},L]}
\end{eqnarray}
where the subscript $+$ denotes the purely differential part: i.e., for a
operator (2.2), $O_{+} = \sum^{N}_{s=0}o_{s}D^{s}$, whereas
$O_{-}$ means the purely pseudo-differential part
$\sum^{-1}_{s=-\infty}o_{s}D^{s}$.
It is easy to verify the equations in the KP hierarchy are compatible,
namely different time evolutions of $L$ commute with each other:
\begin{eqnarray}
\frac{\partial^{2}L}{\partial t_{m}\partial t_{n}} =
\frac{\partial^{2}L}{\partial t_{n}\partial t_{m}}.
\end{eqnarray}
Note that eq.(2.4) implies
\begin{eqnarray}
\frac{\partial L^{N}}{\partial t_{m}} = {[(L^{m})_{+},L^{N}]}
\end{eqnarray}
for any positive integer $N$. Imposing the constraint
$(L^{N})_{-} = 0$, eq.(2.6) becomes a set of differential equations
in terms of an order-$N$ differential KdV operator
$Q \equiv L^N = D^{N} + \sum^{N-1}_{i=0}q_{i}D^{i}$:
\begin{eqnarray}
\frac{\partial Q}{\partial t_{m}} = {[(Q^{m/N})_{+},Q]}
\end{eqnarray}
where $Q^{m/N} \equiv (Q^{1/N})^{m}$ and $Q^{1/N}$ is
the unique pseudo-differential
operator that satisfies $(Q^{1/N})^N$ $= Q$. Hence the KP hierarchy (2.4)
contains all the $N$th generalized KdV hierarchies (2.7) by reduction.

The most interesting feature of the KP hierarchy is its integrability.
It is a direct consequence of its bi-Hamiltonian
structure first proposed by Dickey [22]. To consider the KP
hierarchy from the Hamiltonian point of view, one starts with trying
to put eq.(2.4) into the Hamiltonian form
\begin{eqnarray}
\frac{\partial L}{\partial t_{m}} = K\frac{\delta H_{m}}{\delta u},
\end{eqnarray}
which is equivalent to rewriting the coefficients of
\begin{eqnarray}
{[(L^{m})_{+},L]} = \sum_{r=-1}^{\infty} K_{r}(m)D^{-r-1}
\end{eqnarray}
as
\begin{eqnarray}
K_{r}(m) = \sum_{s=-1}^{\infty} k_{rs} \frac{\delta H_{m}}{\delta u_{s}}
\end{eqnarray}
where $\delta / \delta u_{s}$ stands for the usual variational derivative
\begin{eqnarray}
\frac{\delta}{\delta u_{s}} = \sum_{k=0}
(- \partial_z)^k  \frac{\partial}{\partial  u_{s}^{(k)}}.
\end{eqnarray}
The infinite dimensional operator matrix $k_{rs}$ in eq.(2.10) gives rise to
a Hamiltonian structure if the associated Poisson brackets
\begin{eqnarray}
{\{u_{r}(z),u_{s}(z')\}} = k_{rs}(z) \delta (z-z')
\end{eqnarray}
form a closed algebra. Correspondingly, $H_{m}$ in eq.(2.10) constitute
an infinite set of Hamiltonian functions in this Hamiltonian structure.
Using eq.(2.12), we are able to express eq.(2.8) in the desired
canonical form:
\begin{eqnarray}
\frac{\partial u_{r}}{\partial t_{m}} = K_{r}(m) =
{\{u_{r}(z), \oint H_{m}(z)dz\}}
\end{eqnarray}
with appropriate boundary conditions for the integrands.

In search of Hamiltonian structures, one expands $L^{m}$ as
\begin{eqnarray}
L^{m} = D^{m} + \sum^{m-1}_{s=-\infty}D^{s}U_{s}(m).
\end{eqnarray}
It follows that
\begin{eqnarray}
U_{s}(m) = \frac{1}{m+1} \frac{\delta ResL^{m+1}}{\delta u_{s}}
\end{eqnarray}
for $s\geq -1$. Here the residue, $Res O$, of a pseudo-differential operator
$O$ means the coefficient of the $D^{-1}$ term in $O$. A simple calculation
of the commutator $[(L^{m})_{+},L]$ in eq.(2.9) leads to the first
Hamiltonian form of the KP hierarchy
\begin{eqnarray}
K_{r}^{(1)}(m) &=& \frac{1}{m+1}\sum^{\infty}_{s=-1}k_{rs}^{(1)}
\frac{\delta ResL^{m+1}}{\delta u_{s}} \nonumber\\
&=& {\{u_{r}, \frac{1}{m+1}\oint ResL^{m+1}(z)dz\}}_{1}
\end{eqnarray}
where $k_{rs}^{(1)}$ is the Watanabe (or the first) Hamiltonian structure
[21] and $(m+1)^{-1} ResL^{m+1}$ are corresponding Hamiltonian functions
$H_{m}^{(1)}$. (The script 1 labels the first Hamiltonian structure,
and similarly 2 will label the second one below.)

To obtain the second KP Hamiltonian form, let us express
$[(L^{m})_{+},L]$ as
\begin{eqnarray}
{[(L^{m})_{+},L]} = (LL^{m-1})_{+}L - L(L^{m-1}L)_{+}.
\end{eqnarray}
By substituting (2.14-15) for $L^{m-1}$ into the right hand side,
we can recast it into the form of eq.(2.9) with
\begin{eqnarray}
K_{r}^{(2)}(m) &=& \frac{1}{m}\sum^{\infty}_{s=-1} k_{rs}^{(2)}
\frac{\delta ResL^{m}}{\delta u_{s}} \nonumber\\
&=& {\{u_{r}, \frac{1}{m}\oint ResL^{m}(z)dz\}}_{2}
\end{eqnarray}
where $k_{rs}^{(2)}$ is the second Hamiltonian structure of Dickey [22]
with $(1/m)ResL^{m}$ the associated Hamiltonian functions $H_{m}^{(2)}$.

These two KP hamiltonian structures are compactible in the
sense that they can be linearly combined into a one-parameter
family of Hamiltonian structures, called a bi-Hamiltonian structure
[22]. To see this, for an arbitrary pseudo-differential operator $P$,
we define a ``Hamiltonian'' operator
\begin{eqnarray}
K(P) = (LP)_{+}L-L(PL)_{+}.
\end{eqnarray}
Note $K(L^{m-1})$ will give us the second Hamiltonian form (2.18).
Shift $L$ by a constant:
\begin{eqnarray}
\hat{L}\equiv L+c = D+\sum^{\infty}_{r=-1}\hat{u}_{r}D^{-r-1},
\end{eqnarray}
and define the corresponding
\begin{eqnarray}
\hat{K}(P) = (\hat{L}P)_{+}\hat{L}-\hat{L}(P\hat{L})_{+}.
\end{eqnarray}
To obtain a one-parametric family of KP Hamiltonian
structures labeled by $c$,
let us substitute an operator $P$ of order $N$ into eq.(2.21)
\begin{eqnarray}
P = \sum_{s=-\infty}^{N} D^{s}p_{s}
\end{eqnarray}
($N$ can be arbitrarily large) and rewrite (2.21) as
\begin{eqnarray}
\hat{K}(P) = \sum_{r,s=-1}^{\infty}\hat{k}_{rs}p_{s}D^{-r-1}.
\end{eqnarray}
In the following we are going to show that for fixed pair $(r,s)$,
the operator $\hat{k}_{rs}$ becomes stable (i.e. independent of $N$)
when $N$ is sufficiently large. Thus, we can use these stable
``entries'' to form an infinite-dimensional operator matrix
$\hat{k}_{rs}$, and use it to define the Poisson brackets underlying
the bi-Hamiltonian structure for functions $u_r$:
\begin{eqnarray}
{\{u_{r}(z),u_{s}(z')\}} = \hat{k}_{rs}(\hat{u}_p(z))~
\delta (z-z').
\end{eqnarray}
with $\hat{u}_{-1}=u_{-1}+c$ and $\hat{u}_r=u_r$
($r\geq 0$) in the right side. It is easy to verify the
skew-adjointness of $\hat{k}_{rs}$; the closure
of $\hat{k}_{rs}$ was shown in ref. [22]. We will show that
in the $c\rightarrow 0$ and $c\rightarrow\infty$ limit,
one recovers the second and first KP Hamiltonian structure respectively.

After this general description, now let us derive the explicit expressions
of the KP Hamiltonian structures in closed form. We start with evaluating
the $\hat{k}_{rs}$ for the bi-Hamiltonian structure defined by eq.(2.24).

Proposition 1: For $r,s\geq 0$,
\begin{eqnarray}
\hat{k}_{rs} &=& \sum^{s+1}_{l=0} \left( \begin{array}{c}
s+1 \\ l
\end{array} \right) D^{l}\hat{u}_{r+s+1-l}
-\sum^{r+1}_{l=0} \left( \begin{array}{c}
r+1 \\ l
\end{array} \right) \hat{u}_{r+s+1-l}(-D)^{l} \nonumber\\
& & +\sum^{\infty}_{l=0}{[
\sum^{l+s}_{t=l}(-1)^{l} \left( \begin{array}{c}
t-s-1 \\ l
\end{array} \right)} \nonumber\\
& & {-\sum^{r+s+1}_{t=r+1}\sum^{t-r-1}_{k=0}(-1)^{l+k} \left( \begin{array}{c}
t-k-1 \\ l-k
\end{array} \right) \left( \begin{array}{c}
s \\ k
\end{array} \right) ]}\hat{u}_{t-l-1}D^{l}\hat{u}_{r+s-t}, \nonumber\\
\hat{k}_{r-1} &=& -\sum^{r}_{l=1} \left( \begin{array}{c}
r \\ l
\end{array} \right) \hat{u}_{r-l}(-D)^{l}, \nonumber\\
\hat{k}_{-1s} &=& \sum^{s}_{l=1} \left( \begin{array}{c}
s \\ l
\end{array} \right) D^{l}\hat{u}_{s-l}, ~~~~~ \hat{k}_{-1-1}~=~ -D.
\end{eqnarray}

Proof: First we observe that for a pseudo-differential operator $P$ of
order $N$, eq.(2.21) involves only the following part of it:
$P_{\tilde{+}}\equiv\sum^{\infty}_{s=0}D^{s}p_{s}+p_{-1}D^{-1}$.
Here $p_{s}$ $(s>N)$ are formally set to zero. The advantage of this
convention is that we can take $N$ to be arbitrarily large without
changing the resulting formula. Then obviously we have
\begin{eqnarray}
\hat{K}(P) = (\hat{L}P_{\tilde{+}})_{+}\hat{L}-
\hat{L}(P_{\tilde{+}}\hat{L})_{+} = \hat{L}(P_{\tilde{+}}\hat{L})_{-}
-(\hat{L}P_{\tilde{+}})_{-}\hat{L}.
\end{eqnarray}
Substituting the explicit expressions of $\hat{L}$ and $P_{\tilde{+}}$ into
eq.(2.26), it becomes
\begin{eqnarray}
\hat{K}(P) &=& (D+\sum^{\infty}_{r=-1}\hat{u}_{r}D^{-r-1})
(\sum^{\infty}_{s=0}\sum^{\infty}_{t=-1}D^{s}p_{s}\hat{u}_{t}D^{-t-1}
+\sum^{\infty}_{t=-1}p_{-1}D^{-1}\hat{u}_{t}D^{-t-1})_{-} \nonumber\\
& & -(\sum^{\infty}_{r=-1}\sum^{\infty}_{s=0}\hat{u}_{r}D^{-r-1+s}p_{s}
+\sum^{\infty}_{r=-1}\hat{u}_{r}D^{-r-1}p_{-1}D^{-1})_{-}
(D+\sum^{\infty}_{t=-1}\hat{u}_{t}D^{-t-1}) \nonumber\\
& & -p_{-1}'D^{-1}(D+\sum^{\infty}_{t=-1}\hat{u}_{t}D^{-t-1}) \nonumber\\
&=& (D+\sum^{\infty}_{r=-1}\hat{u}_{r}D^{-r-1})
\sum^{\infty}_{s=0}\sum^{s}_{k=0}\sum^{\infty}_{t=s-k}\left( \begin{array}{c}
s\\k
\end{array} \right) (\hat{u}_{t}p_{s})^{(k)}D^{s-k-t-1} \nonumber\\
& & -\sum^{\infty}_{s=0}\sum^{\infty}_{r=s}\hat{u}_{r}D^{-r-1+s}p_{s}
(D+\sum^{\infty}_{t=-1}\hat{u}_{t}D^{-t-1}) \nonumber\\
& & +\sum^{\infty}_{t=0}p_{-1}\hat{u}_{t}D^{-t-1} -
\sum^{\infty}_{r=0}\hat{u}_{r}D^{-r-1}p_{-1} - p_{-1}' \nonumber
\end{eqnarray}
where $p_{-1}' \equiv \partial_{z}p_{-1}$ and several terms involving
$\hat{u}_{-1}$ or $p_{-1}$ have been cancelled. After moving every $D$
to most left, it follows that
\begin{eqnarray}
\hat{K}(P) &=& \sum^{\infty}_{s=0}\sum^{s}_{k=0}\sum^{\infty}_{t=s-k}
\left( \begin{array}{c}
s\\k
\end{array} \right) [(\hat{u}_{t}p_{s})^{(k+1)}D^{s-k-t-1}+
(\hat{u}_{t}p_{s})^{(k)}D^{s-k-t}] \nonumber\\
& & -\sum^{\infty}_{s=0}\sum^{\infty}_{r=s}\sum^{\infty}_{l=0}
\left( \begin{array}{c}
-r-1+s\\l
\end{array} \right) \hat{u}_{r}p_{s}^{(l)}D^{-r-l+s}  \nonumber\\
& & +\sum^{\infty}_{s=0}\sum^{\infty}_{r=-1}\sum^{s}_{k=0}
\sum^{\infty}_{t=s-k}\sum^{\infty}_{l=0}
\left( \begin{array}{c}
-r-1\\l
\end{array} \right) \left( \begin{array}{c}
s\\k
\end{array} \right) \hat{u}_{r}(\hat{u}_{t}p_{s})^{(k+l)}D^{-r-l+s-k-t-2}
\nonumber\\
& & -\sum^{\infty}_{s=0}\sum^{\infty}_{r=s}
\sum^{\infty}_{t=-1}\sum^{\infty}_{l=0}
\left( \begin{array}{c}
-r-1+s\\l
\end{array} \right) \hat{u}_{r}(\hat{u}_{t}p_{s})^{(l)}D^{-r-l+s-t-2}
\nonumber\\
& & +\sum^{\infty}_{r=0}\hat{u}_{r}p_{-1}D^{-r-1}
-\sum^{\infty}_{r=0}\sum^{\infty}_{l=0} \left( \begin{array}{c}
-r-1\\l
\end{array} \right) \hat{u}_{r}p_{-1}^{(l)}D^{-r-l-1} -p_{-1}'. \nonumber
\end{eqnarray}
Now we rearrange the ordering and dummy indices of the multiple summations
so as to cast $\hat{K}(P)$ into the form of eq.(2.23). We finally obtain
\begin{eqnarray}
\hat{K}(P) &=& \sum^{\infty}_{r,s=0}\{\sum^{s+1}_{l=0} \left( \begin{array}{c}
s+1 \\ l
\end{array} \right) (\hat{u}_{r+s+1-l}p_{s})^{(l)}
-\sum^{r+1}_{l=0} (-1)^{l}\left( \begin{array}{c}
r+1 \\ l
\end{array} \right) \hat{u}_{r+s+1-l}p_{s}^{(l)} \nonumber\\
& & +\sum^{\infty}_{l=0}{[
\sum^{l+s}_{t=l}(-1)^{l} \left( \begin{array}{c}
t-s-1 \\ l
\end{array} \right)
-\sum^{r+s+1}_{t=r+1}\sum^{t-r-1}_{k=0}(-1)^{l+k} \left( \begin{array}{c}
t-k-1 \\ l-k
\end{array} \right) \left( \begin{array}{c}
s \\ k
\end{array} \right) ]} \nonumber\\
& & ~~\cdot \hat{u}_{t-l-1}(\hat{u}_{r+s-t}p_{s})^{(l)}\} D^{-r-1} \nonumber\\
& & -\sum^{\infty}_{r=0}\sum^{r}_{l=1} \left( \begin{array}{c}
r \\ l
\end{array} \right) \hat{u}_{r-l}p_{-1}^{(l)}D^{-r-1}
+\sum^{\infty}_{s=0}\sum^{s}_{l=1} \left( \begin{array}{c}
s \\ l
\end{array} \right) (\hat{u}_{s-l}p_{s})^{(l)} - p_{-1}'.
\end{eqnarray}
One can easily read eq.(2.25) off. (QED)

We note that eq.(2.25) is, though not manifestly, indeed skew-adjoint
as required by the antisymmetry of the Poisson bracket, and is independent
of $N$.

Proposion 2:
\begin{eqnarray}
(\hat{k}_{rs}(z) + \hat{k}_{sr}(w))\delta (z-w) = 0.
\end{eqnarray}

We skip the proof which is straightforward.

Let us show that the KP bi-Hamiltonian structure (2.24)
incorporates both the first and second Hamiltonian structures.

Proposion 3:
\begin{eqnarray}
{\{u_{r}(z),u_{s}(z')\}}_{2} &=& \lim_{c\rightarrow 0}
\hat{k}_{rs}(z) \delta (z-z')~~~~~~~r,s\geq -1; \\
{\{u_{r}(z),u_{s}(z')\}}_{1} &=& \lim_{c\rightarrow \infty}
(\frac{1}{c}\hat{k}_{rs}(z))\delta (z-z')~~~~~r,s\geq 0.
\end{eqnarray}

Proof: The first limit is expected, if we note that by taking it, eq.(2.21)
coincides with (2.19) which leads to the second Hamiltonian structure.
Therefore, $k_{rs}^{(2)}=\hat{k}_{rs}\mid_{c=0}$.

For the second limit, we have
\begin{eqnarray}
\lim_{c\rightarrow\infty}(\frac{1}{c}\hat{K}(P)) = {[P_{+}, L]}
\end{eqnarray}
The right-hand side with $P=L^{m}$ gives rise to the first
Hamiltonian structure. To obtain the explicit expressions from eq.(2.25),
we use the fact that when $c\rightarrow \infty$, only terms which are
bilinear in $\hat{u}$'s and involve the $\hat{u}_{-1}=u_{-1}+c$ current
survive:
\begin{eqnarray}
\lim_{c\rightarrow \infty}(\frac{1}{c}\hat{k}_{rs})
= \sum^{s}_{l=0} \left( \begin{array}{c}
s \\ l
\end{array} \right) D^{l}u_{r+s-l}
-\sum^{r}_{l=0} \left( \begin{array}{c}
r \\ l
\end{array} \right) u_{r+s-l}(-D)^{l}.
\end{eqnarray}
It is exactly the same as derived directly from eq.(2.31) [1].
(The $u_{-1}$ current becomes decoupled from the algebra.)
(QED)

Therefore, the linear first KP Hamiltonian structure is related to
the nonlinear second KP Hamiltonian structure by contraction.
In ref.[1], we have identified the former with the $W_{1+\infty}$
algebra.  When compared to usual generators of $W_{1+\infty}$ we should
assign spin $r+1$ to the current $u_r$ $(r\geq 0)$. We will call the
second Hamiltonian structure (2.25) (with $c=0$) as the
$\hat{W}_{1+\infty}$ algebra, in which the current $u_r$ is considered
to have spin $r+2$ $(r\geq -1)$. Despite the shift in spin, both
algebras have the same spectrum for the
spin of the currents: $s\geq 1$, since $u_{-1}$
decouples and is not included in $W_{1+\infty}$.

It has been convenient to keep $u_{-1}\neq 0$ in the
discussion of the KP bi-Hamiltonian structure. However, from the
dynamical point of view, $u_{-1}$ is trivial. It does not evolve
at all in accordance to  the KP hierarchy (2.4):
$\partial u_{-1}/\partial t_{m} = 0$. Furthermore,
consider the action of $L$ on the space of functions:
$(D+\sum^{\infty}_{r=-1}u_{r}D^{-r-1})(z)f(z)$. With a change in function
$f(z)\rightarrow exp(-\int^{z}u_{-1}(z')dz')f(z)$, one can always
remove $u_{-1}$.  Hence, without loss of generality, we may deal
with the case with  $u_{-1}=0$. In this case, there appears no
modification to the first Hamiltonian structure, as $u_{-1}$ decouples
from $W_{1+\infty}$. Nevertheless, $u_{-1}=0$ is a second class
constraint in $\hat{W}_{1+\infty}$. One needs
the Dirac brackets to handle it. This results in
\begin{eqnarray}
& & {\{u_{r}(z), u_{s}(z')\}}_{2D} ~\equiv ~k_{rs}^{(2D)}(z)\delta (z-z') \\
&=& {\{u_{r}(z), u_{s}(z')\}}_{2}  - \oint\oint dwdw'
{\{u_{r}(z), u_{-1}(w)\}}_{2}{\{u_{-1}(w), u_{-1}(w')\}}_{2}^{-1}
{\{u_{-1}(w'), u_{s}(z')\}}_{2}. \nonumber
\end{eqnarray}

Alternatively we may consider the modification in  the second KP
Hamiltonian structure induced by $u_{-1}=0$.  Eqs.(2.8)-(2.10) become
\begin{eqnarray}
\frac{\partial L}{\partial t_{m}} &=&
(LL^{m-1})_{+}L-L(L^{m-1}L)_{+} \nonumber\\
&=& \frac{1}{m}\sum^{\infty}_{r=-1}
\sum^{\infty}_{s=0} k_{rs}^{(2)}
\frac{\delta ResL^{m}}{\delta u_{s}}D^{-r-1}
 +{[ResL^{m-1},L]}.
\end{eqnarray}
The $D^0$ term in the left side vanishes, so does that
in the right side.  This gives
\begin{eqnarray}
(ResL^{m-1})' = \frac{1}{m}\sum^{\infty}_{s=0} k_{-1s}^{(2)}
\frac{\delta ResL^{m}}{\delta u_{s}}.
\end{eqnarray}
It allows us to express eq.(2.34) in the modified second Hamiltonian form
\begin{eqnarray}
K_{r}^{(2D)}(m) = \frac{1}{m}
\sum^{\infty}_{s=0}k_{rs}^{(2D)}
\frac{\delta ResL^{m}}{\delta u_{s}}~~~~~r\geq 0
\end{eqnarray}
with
\begin{eqnarray}
k_{rs}^{(2D)} = k_{rs}^{(2)} - \sum^{r}_{l=1}(-1)^{l} \left( \begin{array}{c}
r \\ l
\end{array} \right) u_{r-l}D^{l-1}k_{-1s}^{(2)}.
\end{eqnarray}

Proposition 4:  Eq.(2.37) is equivalent to (2.33).

Proof: By definition, ${\{u_{r}(z), u_{s}(z')\}}_{2D}=
k_{rs}^{(2D)}(z)\delta (z-z')$. We need to show
\begin{eqnarray}
k_{rs}^{(2D)} = k_{rs}^{(2)} - k_{r-1}^{(2)}k_{-1-1}^{(2)~-1}k_{-1s}^{(2)}.
\end{eqnarray}
{}From eq.(2.25), $k_{-1-1}^{(2)}=-D$, $k_{r-1}^{(2)}
=-\sum^{r}_{l=1}(-1)^{l} \left( \begin{array}{c}
r\\l
\end{array} \right) u_{r-l}D^{l}$. Eq.(2.38) is therefore identical to (2.37).
(QED)

Explicitly for the modified second Hamiltonian structure we have

Proposition 5:
\begin{eqnarray}
k_{rs}^{(2D)} &=& \sum^{s+1}_{l=0} \left( \begin{array}{c}
s+1 \\ l
\end{array} \right) D^{l}u_{r+s+1-l} -\sum^{r+1}_{l=0} \left( \begin{array}{c}
r+1 \\ l
\end{array} \right) u_{r+s+1-l}(-D)^{l} \nonumber\\
& & +\sum^{\infty}_{l=0}{[
\sum^{l+s}_{t=l+1}(-1)^{l} \left( \begin{array}{c}
t-s-1 \\ l
\end{array} \right)} \nonumber\\
& & { - \sum^{r+s}_{t=r+1}\sum^{t-r-1}_{k=0}(-1)^{l+k} \left( \begin{array}{c}
t-k-1 \\ l-k
\end{array} \right) \left( \begin{array}{c}
s \\ k
\end{array} \right) ]}u_{t-l-1}D^{l}u_{r+s-t} \nonumber\\
& & -\sum^{r}_{l=1}\sum^{s}_{k=1}(-1)^{l}\left( \begin{array}{c}
r\\l
\end{array} \right) \left( \begin{array}{c}
s\\k
\end{array} \right) u_{r-l}D^{l+k-1}u_{s-k}.
\end{eqnarray}

Proof: The first four terms of eq.(2.39) rewrite $k_{rs}^{(2)}$ with
$r,s\geq 0$ and $u_{-1}=0$. The last term is obtained by substituting
$k_{-1s}^{(2)}=\sum^{s}_{k=1}\left( \begin{array}{c}
s\\k
\end{array} \right) D^{k}u_{s-k}$ into eq.(2.37). (QED)

As shown in ref.[2], the modified second KP Hamiltonian structure (2.33)
can be viewed as the unique nonlinear, centerless deformation of
$W_{\infty}$ (and $w_{\infty}$) under certain natural homogeneity
requirements. Therefore we called it as the $\hat{W}_{\infty}$ algebra.
It is a nonlinear current algebra formed by all currents of spin
$s\geq 2$. In this way we have seen that the $\hat{W}_{1+\infty}$
is a universal $W$-algebra, in the sense that it contains all known
$W$-algebras of the infinite type either by contraction or by
reduction. Also it is conjectured [2, 23] to contain all $W_N$-algebras
of the finite type by reduction: Since $W_N$ is known to be isomorphic
to the second Hamiltonian structure of the $N$-th generalized
KdV hierarchy (also modified by a condition similar to $u_{-1}=0$), and
the KP hierarchy contains all generalized KdV hierarchies, it
is natural to expect that the nonlinear $\hat{W}_{\infty}$ reduces
to $W_{N}$ by imposing the second class constraints $(L^N)_{-}=0$.
(Technically it would not be easy to carry out this reduction
because of the infinite number of constraints.) In view of this,
$\hat{W}_{\infty}$ seems physically more interesting than
$\hat{W}_{1+\infty}$. In the following we will focus on the former,
and will suppress the scripts ``$2D$'' for the Poisson brackets
associated with it when no confusion arises.

We note that with $u_{-1}=0$, there is an interesting relation
between $\hat{W}_{\infty}$ and  $W_{1+\infty}$ as follows [2]:
\begin{eqnarray}
\oint {\{\hat{W}_{r+1}(z),\hat{W}_{s+1}(z')\}}_{2D}dz'
= \oint {\{W_{r}(z),W_{s}(z')\}}_{1}dz',
\end{eqnarray}
where the $\hat{W}_{\infty}$-current $\hat{W}_{r+1}$ and
$W_{1+\infty}$-current
$W_{r}$ $(r\geq 1)$ are respectively the Hamiltonian functions
\begin{eqnarray}
\hat{W}_{r+1} \equiv H_{r}^{(2D)} = \frac{1}{r}ResL^{r};
{}~~~W_{r} \equiv H_{r-1}^{(1)} = \frac{1}{r}ResL^{r}.
\end{eqnarray}
This relation is a consequence of the coexistence of two KP
Hamiltonian structures, which implies the generalized
Lenard recursion relation in the KP hierarchy:
\begin{eqnarray}
\frac{\partial u_{r}}{\partial t_{m}} = K_{r}^{(2D)}(m) = K_{r}^{(1)}(m),
\end{eqnarray}
or canonically, from eq.(2.13),
\begin{eqnarray}
{\{u_{r},\oint H_{m}^{(2D)}(z)dz\}}_{2D}
= {\{u_{r},\oint H_{m}^{(1)}(z)dz\}}_{1}.
\end{eqnarray}
By multiplying both sides of (2.43) by $\delta \hat{W}_{s+1}/\delta u_{r}
(=\delta W_{s}/\delta u_{r})$, it is clear that (2.43) is identical to
(2.40).

In conclusion of this section, we mention that the supersymmetric version of
KP bi-Hamiltonian structure thus the complete integrability of a super KP
hierarchy, and its relation to the super $W$-algebras
of the infinite type have been worked out in refs.[25,26]

\vspace{30 pt}
\section{Free Field Realization of $\hat{W}_{\infty}$}
\setcounter{equation}{0}
\vspace{5 pt}

The main issue of this section is to construct a two boson
realization of the nonlinear $\hat{W}_{\infty}$ algebra in closed form,
which will play a crucial role in our later discussions.

To begin with, let us briefly review some known results of free field
representations of $W$-algebras of both finite and infinite type,
in terms of (pseudo-)differential operators. We will make extensive
use of the techniques in operator calculus [15]. First,
$W_{1+\infty}$ can be realized with a pair of fermions [13]
$\bar{\psi}(z)$ and $\psi (z)$ satisfying the Poisson brackets
\begin{eqnarray}
{\{\bar{\psi}(z), \psi (z')\}} &=& \delta (z-z'), \nonumber\\
{\{\bar{\psi}(z), \bar{\psi} (z')\}} &=& {\{\psi (z), \psi (z')\}} ~=~ 0.
\end{eqnarray}
In the KP basis [1], the $W_{1+\infty}$ generators $u_{r}(z)$ are simply
summarized by
\begin{eqnarray}
L\equiv D+\sum^{\infty}_{r=0}u_{r}D^{-r-1}=D-\bar{\psi}D^{-1}\psi.
\end{eqnarray}
The Poisson brackets among $u_{r}$ from eq.(3.1) exactly yield
eq.(2.30) plus (2.32). Now considering only the spin $s\geq 2$ currents in
eq.(3.2), the $W_{\infty}$ generators $v_{r}$ naturally have a free fermion
realization too:
\begin{eqnarray}
(LD)_{-}\equiv\sum^{\infty}_{r=0}v_{r}D^{-r-1}
=-(\bar{\psi}D^{-1}\psi D)_{-}.
\end{eqnarray}
Meanwhile, $W_{\infty}$ can be realized in terms of a pair of free bosons [14]
$\bar{\phi}(z)$ and $\phi (z)$, with their currents
$\bar{j}(z)=\bar{\phi}'(z)$, $j(z)=\phi'(z)$ satisfying
\begin{eqnarray}
{\{\bar{j}(z), j(z')\}} &=& \partial_{z}\delta(z-z'); \nonumber\\
{\{\bar{j}(z), \bar{j}(z')\}} &=& {\{j(z), j(z')\}} ~=~ 0.
\end{eqnarray}
The expression of $v_{r}(z)$ is given by
\begin{eqnarray}
L\equiv D+\sum^{\infty}_{r=0}v_{r}D^{-r-1}=D+\bar{j}D^{-1}j.
\end{eqnarray}
The Poisson brackets among $v_{r}$ either from eqs.(3.4)-(3.5) or from
(3.1) and (3.3) precisely lead to those of $W_{\infty}$,
or the linear part of eq.(2.33) plus (2.39).

On the other hand, recall that for finite $N$, the differential KdV
operator of order $N$ can be expressed via the Miura transformation
[27] in terms of $N$ free bosons $\phi_{i}$:
\begin{eqnarray}
Q = D^{N} + \sum^{N-1}_{i=0}q_{i}D^{i} = \prod^{N}_{i=1}(D+j_{i})
\end{eqnarray}
where $j_{i}=\phi_{i}'$.
The second Hamiltonian structure of the $N$-th generalized KdV
hierarchy (2.7) with $q_{i}$ in eq.(3.6) as generators then has a free boson
realization through the Poisson brackets among $j_{i}$:
\begin{eqnarray}
{\{j_{i}(z), j_{k}(z')\}} = \delta_{i,k}\partial_{z}\delta(z-z').
\end{eqnarray}
The simple reduction of this Hamiltonian structure with the
(second class) constraint $q_{N-1}=0$ gives
rise to the $N-1$ free boson representation of the
$W_{N}$ algebra.

In turn, for the desired $\hat{W}_{\infty}$ case, we notice the fact that
the nonlinearities of $\hat{W}_{\infty}$ emerge as deformation of $W_{\infty}$,
which is nicely realized by eq.(3.5) with only two free bosons. This motivates
us to search for a free field realization of $\hat{W}_{\infty}$ in terms
of two scalars by adding higher nonlinear terms to (3.5).
Our main result is summarized in the following KP operator
(with $u_{-1}=0$)
\begin{eqnarray}
L &=& D+\sum^{\infty}_{r=0}u_{r}D^{-r-1}
= D+\bar{j}\frac{1}{D-(\bar{j}+j)}j \nonumber\\
&=& D+\bar{j}D^{-1}j+\bar{j}D^{-1}(\bar{j}+j)D^{-1}j+ \cdots .
\end{eqnarray}
written in terms of the chiral currents $j(z)$ and $\bar{j}(z)$.

In this way, the $\hat{W}_{\infty}$ generators $u_{r}$ are expressed
as functions of $\bar{j}$ and $j$.
Notice the first two terms, $D+\bar{j}D^{-1}j$, realizes the
linear $W_{\infty}$ in the KP basis (3.5),
while the remaining terms in $\bar{j}$ and $j$ represent
the nonlinear deformation of $\hat{W}_{\infty}$. Explicitly,
the first a few $u_{r}$ have the following expression:
\begin{eqnarray}
u_{0} &=& \bar{j}j, \nonumber\\
u_{1} &=& -\bar{j}j'+\bar{j}j^{2}+\bar{j}^{2}j, \nonumber\\
u_{2} &=& \bar{j}j''-3\bar{j}jj'-2\bar{j}^{2}j'-\bar{j}\bar{j}'j
+\bar{j}j^{3}+2\bar{j}^{2}j^{2}+\bar{j}^{3}j, \nonumber\\
u_{3} &=& -\bar{j}j'''+4\bar{j}jj''+3\bar{j}j'^{2}+3\bar{j}^{2}j''
+3\bar{j}\bar{j}'j'+\bar{j}\bar{j}''j \nonumber\\
& & -6\bar{j}j^{2}j'-9\bar{j}^{2}jj'-3\bar{j}\bar{j}'j^{2}-3\bar{j}^{3}j'
-3\bar{j}^{2}\bar{j}'j \nonumber\\
& & +\bar{j}j^{4}+3\bar{j}^{2}j^{3}+3\bar{j}^{3}j^{2}+\bar{j}^{4}j.
\end{eqnarray}
By checking the Poisson brackets among them, we find eq.(3.9) is in fact the
unique two boson realization of these $\hat{W}_{\infty}$ generators up to
isomorphisms. Nevertheless the key issue is to prove the
realization (3.9) for all generators $u_{r}$. This amounts to showing
that the Poisson brackets among $u_{r}$, evaluated according to eq.(3.4),
not only form a closed algebra, but also are identical to
those of $\hat{W}_{\infty}$ (2.33), the modified second
Hamiltonian structure of the KP hierarchy.
Using operator calculus, this is equivalent to proving the
brackets between two functionals
$\oint f(u_{r}(z))dz$ and $\oint g(u_{s}(z))dz$ satisfy
\begin{eqnarray}
{\{\oint f(u_{r}(z))dz, \oint g(u_{s}(z'))dz'\}}
= \oint \sum^{\infty}_{r,s=0}\frac{\delta f}{\delta u_{r}}k_{rs}^{(2D)}
\frac{\delta g}{\delta u_{s}} (z)dz.
\end{eqnarray}

First, the variation of the functional $\oint fdz$ of $u_{r}$ and their
derivatives is given by
\begin{eqnarray}
\delta\oint fdz = \sum^{\infty}_{r=0}\oint\delta u_{r}
\frac{\delta f}{\delta u_{r}}dz = \oint (\delta\bar{j}\frac{\delta f}{\delta
\bar{j}}+ \delta j\frac{\delta f}{\delta j})dz
\end{eqnarray}
where $\delta f/\delta u_{r}$, etc, are the usual variational derivatives
defined by eq.(2.11).
Let us introduce the variational derivative with respect to the KP operator
$L$ to summarize the variations $\delta f/\delta u_{r}$:
\begin{eqnarray}
\frac{\delta f}{\delta L} \equiv \sum^{\infty}_{r=0} D^{r}
\frac{\delta f}{\delta u_{r}}.
\end{eqnarray}
Then eq.(3.11) can be put into a neat form
\begin{eqnarray}
\delta\oint fdz = \oint Res(\delta L \frac{\delta f}{\delta L})dz
\end{eqnarray}
in which
\begin{eqnarray}
\delta L &=& \sum^{\infty}_{r=0}\delta u_{r} D^{-r-1} \nonumber\\
&=& (1+\bar{j}\frac{1}{D-(\bar{j}+j)})\delta\bar{j}\frac{1}{D-(\bar{j}+j)}j
\nonumber\\
& & +\bar{j}\frac{1}{D-(\bar{j}+j)}\delta j(1+\frac{1}{D-(\bar{j}+j)}j).
\end{eqnarray}
Therefore we obtain the following expressions of
$\delta f/\delta\bar{j}$ and $\delta f/\delta j$ by comparing eqs.(3.11)
with (3.13):
\begin{eqnarray}
\frac{\delta f}{\delta \bar{j}} &=& Res(\frac{1}{D-(\bar{j}+j)}j
\frac{\delta f}{\delta L}(1+\bar{j}\frac{1}{D-(\bar{j}+j)})), \nonumber\\
\frac{\delta f}{\delta j} &=& Res((1+\frac{1}{D-(\bar{j}+j)}j)
\frac{\delta f}{\delta L}\bar{j}\frac{1}{D-(\bar{j}+j)}).
\end{eqnarray}
Now we proceed to evaluate the Poisson bracket
\begin{eqnarray}
{\{\oint fdz, \oint gdz' \}} = \oint (\frac{\delta f}{\delta \bar{j}}
(\frac{\delta g}{\delta j})' + \frac{\delta f}{\delta j}
(\frac{\delta g}{\delta \bar{j}})')dz.
\end{eqnarray}
Replacing $(\delta g/\delta j)'$ by the commutator ${[D-(\bar{j}+j),
\delta g/\delta j ]}$, similarly for $(\delta g/\delta \bar{j})'$, and
substituting eq.(3.15) for $\delta f/\delta j$, $\delta g/\delta j$, etc.,
into eq.(3.16), we have
\begin{eqnarray}
{\{\oint fdz, \oint gdz' \}}
&=& \oint Res\{\frac{1}{D-(\bar{j}+j)}j\frac{\delta f}{\delta L}
(1+\bar{j}\frac{1}{D-(\bar{j}+j)}) \nonumber\\
& & ~~~[(D-(\bar{j}+j))Res(
(1+\frac{1}{D-(\bar{j}+j)}j)\frac{\delta g}{\delta L}
\bar{j}\frac{1}{D-(\bar{j}+j)}) \nonumber\\
& & ~~~-Res((1+\frac{1}{D-(\bar{j}+j)}j)\frac{\delta g}{\delta L}
\bar{j}\frac{1}{D-(\bar{j}+j)})(D-(\bar{j}+j))] \nonumber\\
& & ~~~ +(1+\frac{1}{D-(\bar{j}+j)}j)\frac{\delta f}{\delta L}
\bar{j}\frac{1}{D-(\bar{j}+j)} \\
& & ~~~[(D-(\bar{j}+j))Res(\frac{1}{D-(\bar{j}+j)}j
\frac{\delta g}{\delta L} (1+\bar{j}\frac{1}{D-(\bar{j}+j)})) \nonumber\\
& & ~~~-Res(\frac{1}{D-(\bar{j}+j)}j
\frac{\delta g}{\delta L} (1+\bar{j}\frac{1}{D-(\bar{j}+j)}))
(D-(\bar{j}+j))]\} dz. \nonumber
\end{eqnarray}
It follows, by using the identities $ResP=(P_{-}
(D-(\bar{j}+j)))_{+}=((D-(\bar{j}+j))P_{-})_{+}$,
\begin{eqnarray}
& & {\{\oint fdz, \oint gdz' \}} \nonumber\\
&=& \oint Res\{\frac{1}{D-(\bar{j}+j)}j\frac{\delta f}{\delta L}
(1+\bar{j}\frac{1}{D-(\bar{j}+j)}) \nonumber\\
& & ~~~[(D-(\bar{j}+j))
((1+\frac{1}{D-(\bar{j}+j)}j)\frac{\delta g}{\delta L}
\bar{j}\frac{1}{D-(\bar{j}+j)}(D-(\bar{j}+j)))_{+} \nonumber\\
& & ~~~ -((D-(\bar{j}+j))(1+\frac{1}{D-(\bar{j}+j)}j)
\frac{\delta g}{\delta L}\bar{j}\frac{1}{D-(\bar{j}+j)})_{+}(D-(\bar{j}+j))]
\nonumber\\
& & ~~~ +(1+\frac{1}{D-(\bar{j}+j)}j)\frac{\delta f}{\delta L}
\bar{j}\frac{1}{D-(\bar{j}+j)} \\
& & ~~~[(D-(\bar{j}+j))(\frac{1}{D-(\bar{j}+j)}j
\frac{\delta g}{\delta L} (1+\bar{j}\frac{1}{D-(\bar{j}+j)})
(D-(\bar{j}+j)))_{+} \nonumber\\
& & ~~~ -((D-(\bar{j}+j))\frac{1}{D-(\bar{j}+j)}j
\frac{\delta g}{\delta L} (1+\bar{j}\frac{1}{D-(\bar{j}+j)}))_{+}
(D-(\bar{j}+j))]\} dz. \nonumber
\end{eqnarray}
A direct calculation gives
\begin{eqnarray}
{\{\oint fdz, \oint gdz' \}} &=& \oint Res[\bar{j}\frac{1}{D-(\bar{j}+j)}j
\frac{\delta f}{\delta L}(D-j)((1+\frac{1}{D-(\bar{j}+j)}j)
\frac{\delta g}{\delta L})_{+} \nonumber\\
& & ~~~-\frac{\delta f}{\delta L}(1+\bar{j}\frac{1}{D-(\bar{j}+j)})
((D-j)\frac{\delta g}{\delta L}\bar{j}\frac{1}{D-(\bar{j}+j)}j)_{+} \nonumber\\
& & ~~~ +(1+\frac{1}{D-(\bar{j}+j)}j)\frac{\delta f}{\delta L}
(\bar{j}\frac{1}{D-(\bar{j}+j)}j\frac{\delta g}{\delta L}(D-j))_{+} \nonumber\\
& & ~~~ -(D-j)\frac{\delta f}{\delta L}\bar{j}\frac{1}{D-(\bar{j}+j)}j
(\frac{\delta g}{\delta L}(1+\bar{j}\frac{1}{D-(\bar{j}+j)}))_{+} ]dz
\nonumber\\
&=& \oint Res[\frac{\delta f}{\delta L}(L\frac{\delta g}{\delta L})_{+}L
-\frac{\delta f}{\delta L}L(\frac{\delta g}{\delta L}L)_{+}
+ \frac{\delta f}{\delta L}((L-D)\frac{\delta g}{\delta L})_{+}L \nonumber\\
& & ~~~-\frac{\delta f}{\delta L}L(\frac{\delta g}{\delta L}(L-D))_{+} ]dz
\end{eqnarray}
where we have frequently applied the theorem
\begin{eqnarray}
\oint Res{[P,Q]}dz=0
\end{eqnarray}
for arbitrary two operators $P$ and $Q$.

It remains to show that eq.(3.19) really leads to (3.10). Notice that
the expression $(L\frac{\delta g}{\delta L})_{+}L-
L(\frac{\delta g}{\delta L}L)_{+}$ in the first two terms of (3.19) is
actually the (unmodified) second KP Hamiltonian form (2.19) or
(2.21) with the operator $P$ there being $\delta g/\delta L$. So we need
to identify the last two terms of (3.19) to be the modification induced
by the constraint $u_{-1}=0$.

On one hand, the last two terms of (3.19) are equal to
\begin{eqnarray}
& & \oint Res [\frac{\delta f}{\delta L}( (\bar{j}\frac{1}{D-(\bar{j}+j)}j
\frac{\delta g}{\delta L})_{+}\bar{j}\frac{1}{D-(\bar{j}+j)}j \nonumber\\
& & ~~~-\bar{j}\frac{1}{D-(\bar{j}+j)}j(\frac{\delta g}{\delta L}
\bar{j}\frac{1}{D-(\bar{j}+j)}j)_{+})] dz \nonumber\\
&=& \oint Res [\frac{\delta f}{\delta L}\bar{j}\frac{1}{D-(\bar{j}+j)}
Res(\frac{\delta g}{\delta L}L-L\frac{\delta g}{\delta L})
\frac{1}{D-(\bar{j}+j)}j] dz.
\end{eqnarray}
On the other hand, parallel to the analysis in the last section, if we
denote the unmodified second KP Hamiltonian form as
\begin{eqnarray}
K(P) = (LP)_{+}L-L(PL)_{+}
\end{eqnarray}
where $P_{+}\equiv \delta g/\delta L$ and $P_{-}=(ResP)D^{-1}$, the
corresponding modified Hamiltonian form (with $u_{-1}=0$) appears to be
\begin{eqnarray}
K(P_{+}) &=& ((LP)_{+}L-L(PL)_{+})_{-} \nonumber\\
&=& ((L\frac{\delta g}{\delta L})_{+}L-L(\frac{\delta g}{\delta L}L)_{+}
+{[ResP, L]})_{-}.
\end{eqnarray}
Note that
\begin{eqnarray}
(ResP)' = Res(\frac{\delta g}{\delta L}L-L\frac{\delta g}{\delta L}).
\end{eqnarray}
Moreover, in the realization (3.8), we simply have
\begin{eqnarray}
{[ResP, L]}_{-} &=& \bar{j}(ResP \frac{1}{D-(\bar{j}+j)} -
\frac{1}{D-(\bar{j}+j)}ResP)j \nonumber\\
&=& \bar{j}\frac{1}{D-(\bar{j}+j)}(ResP)'\frac{1}{D-(\bar{j}+j)}j.
\end{eqnarray}
Thus, by substituting eq.(3.24) into (3.25) and comparing it with (3.21),
we obtain
\begin{eqnarray}
{\{\oint fdz, \oint gdz' \}} &=& \oint Res[ \frac{\delta f}{\delta L}
((L\frac{\delta g}{\delta L})_{+}L-L(\frac{\delta g}{\delta L}L)_{+}
+{[ResP, L]}) ]dz \nonumber\\
&=& \oint Res(\frac{\delta f}{\delta L} K(\frac{\delta g}{\delta L}))dz
\end{eqnarray}
as desired.

Finally we come up with

Proposition 6: Eq.(3.10) holds true.

Proof: The modified second Hamiltonian form (3.23) can be rewritten as
(recall $\delta g/\delta L$ is defined by eq.(3.12))
\begin{eqnarray}
K(\frac{\delta g}{\delta L}) = \sum_{r,s=0}^{\infty}k_{rs}
\frac{\delta g}{\delta u_{s}}D^{-r-1}.
\end{eqnarray}
Substituting eqs.(3.27) and (3.12) into (3.26), we have
\begin{eqnarray}
{\{\oint fdz, \oint gdz' \}} &=& \oint Res(\sum^{\infty}_{r,s,t=0}
D^{t}\frac{\delta f}{\delta u_{t}}k_{rs}\frac{\delta g}{\delta u_{s}}
D^{-r-1})dz \nonumber\\
&=& \oint \sum^{\infty}_{r,s=0}
\frac{\delta f}{\delta u_{t}}k_{rs}\frac{\delta g}{\delta u_{s}}dz.
\end{eqnarray}
{}From Propositions 4 and 5 in last section, the infinite
dimensional matrix $k_{rs}$ in eqs.(3.27) and (3.28) is none but the
$\hat{W}_{\infty}$ structure (2.39). (QED)

This proves our two free boson realization of the
$\hat{W}_{\infty}$ algebra (3.8) to all orders.

\vspace{30 pt}
\section{$\hat{W}_{\infty}$ and $SL(2,R)/U(1)$ Coset Model}
\setcounter{equation}{0}
\vspace{5 pt}

As is well known, infinite dimensional algebras, such as the Virasoro
algebra [6], Kac-Moody algebra [28] and extended $W_{N}$ algebras [8],
have shown up in several large classes of 2d conformal field models,
such as minimal models [7], Wess-Zumino-Witten models [29] and
some compact coset models [30], and have played a central role
in solving them. It is natural to explore the connection between
the $\hat{W}_{\infty}$ algebra and more general conformal models.
In this section, we are going to show that $\hat{W}_{\infty}$ indeed
arises in the noncompact $SL(2,R)/U(1)$ coset model [31], and to
present a construction of all $\hat{W}_{\infty}$
currents in closed form, which exploits the known free
boson realization of the coset model [32,33].

Recall the boson realization of the $SL(2,R)_{k}$ current algebra
at the classical level
\begin{eqnarray}
J_{\pm} &=& \sqrt{\frac{k}{2}}e^{\pm\sqrt{\frac{2}{k}}\phi_{3}}
(\phi_{1}' \mp i\phi_{2}')
e^{\pm\sqrt{\frac{2}{k}}\phi_{1}}, \nonumber\\
J_{3} &=& -\sqrt{\frac{k}{2}}\phi_{3}',
\end{eqnarray}
with $\phi_{i}$ denoting three free bosons [32]. When
taking the coset $SL(2,R)_{k}/U(1)$ or gauging $U(1)$,
one considers only vertex operators that commute with the $U(1)$
current $J_{3}$. This is equivalent to imposing the restriction
$J_{3}=0$ or simply $\phi_{3}=0$. Thus we are interested only in the
$J_{\pm}$ part of eq.(4.1), which now depend only on two bosons
$\phi_{1}=(1/\sqrt{2})(\phi+\bar{\phi})$ and
$\phi_{2}=(1/\sqrt{2}i)(\phi-\bar{\phi})$. They become just
the $SL(2,R)_{k}/U(1)$ parafermion currents [33]:
\begin{eqnarray}
\psi_{+}=\bar{j}e^{\bar{\phi}+\phi}, ~~~~
\psi_{-}=je^{-\bar{\phi}-\phi}.
\end{eqnarray}
Here and below, we set the level parameter, $k$, to unity.
This does not lose any generality at the classical level.
To generate higher-spin $\hat{W}_{\infty}$ currents from the
$SL(2,R)/U(1)$ currents (4.2), we propose to study their ordinary
bilocal product $\psi_{+}(z)\psi_{-}(z')$,
and expand this product in powers of $z-z'\equiv \epsilon$ to all orders:
\begin{eqnarray}
\bar{j}e^{\bar{\phi}+\phi}(z)je^{-\bar{\phi}-\phi}(z')
= \sum^{\infty}_{r=0}c_{r}(z)\frac{\epsilon^{r}}{r!}.
\end{eqnarray}
This expansion is the classical counterpart of the operator product
expansion. Our main result is that
the bilocal product (4.3) is a generating function of $\hat{W}_{\infty}$,
in the sense that the coefficient functions $c_{r}(z)$ in
(4.3), as functions of $j$ and $\bar{j}$, are nothing but
the $\hat{W}_{\infty}$ generators $u_r(z)$ realized by eq.(3.8) or (3.9):
\begin{eqnarray}
c_{r}(z)=u_{r}(z).
\end{eqnarray}

To give a compact notation, it is convenient to introduce
the generating function of a pseudo-differential operator
$P(z)=\sum^{\infty}_{r=0}p_{r}(z)D^{-r-1}$ as a
bilocal function $F(z,z')=\sum^{\infty}_{r=0}p_{r}(z)(z-z')^{r}/r!$, and
denote this by $P(z) \Longleftrightarrow F(z,z')$.

Then eq.(4.4) can be restated as

Proposion 7:
\begin{eqnarray}
\bar{j}\frac{1}{D-(\bar{j}+j)}j(z) \Longleftrightarrow
\bar{j}e^{\bar{\phi}+\phi}(z)je^{-\bar{\phi}-\phi}(z').
\end{eqnarray}

Proof: Let us write down the expansion of the right hand side
\begin{eqnarray}
& & \bar{j}e^{\bar{\phi}+\phi}(z)je^{-\bar{\phi}-\phi}(z') \nonumber\\
&=& \bar{j}e^{\bar{\phi}+\phi}\sum^{\infty}_{n=0}\frac{(-1)^{n}}{n!}j^{(n)}
\epsilon^{n}e^{-\sum^{\infty}_{m=0}\frac{(-1)^{m}}{m!}(\bar{\phi}+\phi)^{(m)}
\epsilon^{m}}  \nonumber\\
&=& \sum^{\infty}_{n=0}\sum^{\infty}_{k=0}\frac{(-1)^{n}}{n!k!}\bar{j}j^{(n)}
\epsilon^{n}(\sum^{\infty}_{m=0}\frac{(-1)^{m}}{(m+1)!}(\bar{j}+j)^{(m)}
\epsilon^{m+1})^{k}
\end{eqnarray}
where $\epsilon = z-z'$. As anticipated, eq.(4.6) only depends on
the chiral currents $\bar{j}$ and $j$ (not on $\bar{\phi}$
and $\phi$ themselves). The expansion of the left side of eq.(4.5) is
\begin{eqnarray}
\bar{j}\frac{1}{D-(\bar{j}+j)}j(z) = \sum^{\infty}_{k=0}\bar{j}
(D^{-1}(\bar{j}+j))^{k}D^{-1}j.
\end{eqnarray}
So, by counting the powers in $\bar{j}$ and $j$ in eqs.(4.6) and
(4.7), it is sufficient to show
\begin{eqnarray}
& & \bar{j}(D^{-1}(\bar{j}+j))^{k}D^{-1}j \nonumber\\
&\Longleftrightarrow &
\sum^{\infty}_{n=0}\frac{(-1)^{n}}{n!k!}\bar{j}j^{(n)}
\epsilon^{n}(\sum^{\infty}_{m=0}\frac{(-1)^{m}}{(m+1)!}(\bar{j}+j)^{(m)}
\epsilon^{m+1})^{k}.
\end{eqnarray}
Indeed, by moving all factors $D^{-1}$ in the left hand
side to the most right, we have
\begin{eqnarray}
& & \bar{j}(D^{-1}(\bar{j}+j))^{k}D^{-1}j \nonumber\\
&=& \sum^{\infty}_{m_{1},m_{2},...,m_{k},n=0}
\left( \begin{array}{c}
-1\\m_{1}
\end{array} \right)
\left( \begin{array}{c}
-m_{1}-2\\m_{2}
\end{array} \right) \cdots
\left( \begin{array}{c}
-m_{1}-m_{2}-\cdots -m_{k}-k-1\\n
\end{array} \right) \nonumber\\
& & \bar{j}(\bar{j}+j)^{(m_{1})}(\bar{j}+j)^{(m_{2})}\cdots
(\bar{j}+j)^{(m_{k})}j^{(n)}
D^{-m_{1}-m_{2}-\cdots -m_{k}-n-k-1}.
\end{eqnarray}
By applying the identity $\left( \begin{array}{c}
-a\\b
\end{array} \right) = (-1)^{b} \left( \begin{array}{c}
a+b-1\\b
\end{array} \right) $
and then totally symmetrizing the indices $m_{i}$, it follows that
\begin{eqnarray}
& & \bar{j}(D^{-1}(\bar{j}+j))^{k}D^{-1}j \nonumber\\
&=& \sum^{\infty}_{m_{1},m_{2},...,m_{k},n=0}
(-1)^{m_{1}+m_{2}+\cdots +m_{k}+n}
\frac{(m_{1}+m_{2}+\cdots +m_{k}+n+k)!}{(m_{1}+1)!(m_{2}+1)!\cdots (m_{k}+1)!
n!k!} \nonumber\\
& & \bar{j}(\bar{j}+j)^{(m_{1})}(\bar{j}+j)^{(m_{2})}\cdots
(\bar{j}+j)^{(m_{k})}j^{(n)}D^{-m_{1}-m_{2}
-\cdots -m_{k}-n-k-1} \nonumber\\
&\Longleftrightarrow& \sum^{\infty}_{m_{1},m_{2},...,m_{k},n=0}
\frac{(-1)^{m_{1}+m_{2}+\cdots +m_{k}+n}}{(m_{1}+1)!(m_{2}+1)!
\cdots (m_{k}+1)!n!k!} \nonumber\\
& & \bar{j}(\bar{j}+j)^{(m_{1})}(\bar{j}+j)^{(m_{2})}\cdots
(\bar{j}+j)^{(m_{k})}j^{(n)}
\epsilon^{m_{1}+m_{2}+\cdots +m_{k}+k+n}.
\end{eqnarray}
On the right side all $m_{i}$ sums factorize, and we obtain
the right side of eq.(4.8). (QED)

We can rewrite eqs.(4.3) and (4.4) as
\begin{eqnarray}
u_r(z) = \psi_{+}(z)(-\partial_z)^r \psi_{-}(z),
\end{eqnarray}
or according to the KP operator
\begin{eqnarray}
L = D + \sum^{\infty}_{r=0}u_{r}D^{-r-1} = D + \psi_{+}D^{-1}\psi_{-}.
\end{eqnarray}
It gives us a construction of all the $\hat{W}_{\infty}$ currents (in the
KP basis), in a very compact form, in terms of the basic $SL(2,R)/U(1)$
parafermion or coset currents. Combining Propostions 7 and 6,
these currents must satisfy the $\hat{W}_{\infty}$ algebra, or their
Poisson brackets must be isomorphic to those of the (modified) second
KP Hamiltonian structure.  This is the first time in the literature
that a current algebra with infinitely many independent currents is shown
to exist in a noncompact coset model, in closed form.
It also follows
\begin{eqnarray}
& & {\{\bar{j}e^{\bar{\phi}+\phi}(z)je^{-\bar{\phi}-\phi}(z-\epsilon),
\bar{j}e^{\bar{\phi}+\phi}(w)je^{-\bar{\phi}-\phi}(w-\sigma)\}} \nonumber\\
&\equiv & {\{\psi_{+}(z)\psi_{-}(z-\epsilon),
\psi_{+}(w)\psi_{-}(w-\sigma)\}} \nonumber\\
&=& \sum^{\infty}_{r,s=0} k_{rs}(z) \delta (z-w) \frac{\epsilon^{r}
\sigma^{s}}{r!s!}.
\end{eqnarray}
It is possible to verify this by direct calculation of the left side.

\vspace{30 pt}
\section{Involutive KP charges and $\hat{W}_{\infty}$ Symmetry in the
$SL(2,R)/U(1)$ Coset Model}
\setcounter{equation}{0}
\vspace{5 pt}

We have shown that the $\hat{W}_{\infty}$ algebra appears in both the
integrable KP hierarchy and the conformal $SL(2,R)/U(1)$ model. This
actually implies a close connection between the KP hierarchy and
the coset model. In particular, KP flows are expected to be realizable
by two free boson fields in the coset model.

Let us start with the integrability of the KP hierarchy. First, in the KP
Hamiltonian form (2.13), the Hamiltonian functions
\begin{eqnarray}
H_{m}\equiv H_{m}^{(2)} =H_{m-1}^{(1)} = \frac{1}{m}Res L^{m}
\end{eqnarray}
are conserved charge densities. Namely their integrals are invariant under
the KP flows:
\begin{eqnarray}
\frac{\partial}{\partial t_{n}}\oint H_{m}(z,t_{k}) dz =0.
\end{eqnarray}
Secondly, these infinitely many charges
\begin{eqnarray}
Q_{m}\equiv \oint H_{m}(z) dz
\end{eqnarray}
are in involution - they are independent, commuting charges -
with respect to both the first and second Hamiltonian structures
(regardless of whether $u_{-1}=0$ or $u_{-1}\neq 0$, though we will
present the formulas only for the former case)
\begin{eqnarray}
{\{Q_{n}^{(2)}, Q_{m}^{(2)}\}}_{2} = {\{Q_{n-1}^{(1)}, Q_{m}^{(1)}\}}_{1}
=0.
\end{eqnarray}
This ensures the compatibility and complete integrability of the KP hierarchy
(2.4).

One can prove either of eqs.(5.2) and (5.4) first, then the other follows.
For example, with the substitution
$f=H_{n}$ and $g=H_{m}$ in eq.(3.26), it becomes
\begin{eqnarray}
{\{Q_{n}^{(2)}, Q_{m}^{(2)}\}}_{2} &=&
\oint Res(L^{n-1}K(L^{m-1}))dz \nonumber\\
&=& \oint Res(L^{n-1}({[(L^{m})_{+},L]}+{[ResP , L]})) dz ~=~ 0 \nonumber
\end{eqnarray}
where we have used $\delta ((1/m)ResL^{m})/\delta L = L^{m-1}$
and the identity (3.20). Then the recursion relations (2.40) tells
${\{Q_{n}^{(1)}, Q_{m}^{(1)}\}}_{1} =0$ and eq.(5.2) follows
by applying eq.(2.13).

The charge densities (5.1) are just the $\hat{W}_{\infty}$-currents,
$\hat{W}_{r+1}$, in the Hamiltonian basis (2.41). The existence of the
infinite set of corresponding involutive KP charges (5.3) implies the
existence of as many commuting $\hat{W}_{\infty}$ charges in the
$SL(2,R)/U(1)$ coset model. These $\hat{W}_{\infty}$
charges give rise to a huge infinite
dimensional symmetry in this noncompact conformal model,
which is generated by the multi-time KP flows
\begin{eqnarray}
\frac{\partial u_{r}}{\partial t_{m}} = {\{u_{r}, Q_{m}\}}.
\end{eqnarray}
Or, infinitesimally under a $\hat{W}_{\infty}$
transformation we have
\begin{eqnarray}
\delta_{m} u_{r} = \epsilon_{m} {\{u_{r}, Q_{m}\}}
\end{eqnarray}
with $\epsilon_{m}$ the infinitesimal parameters.
An important property of these symmetry transformations is that
the $\hat{W}_{\infty}$ current algebra (2.33) is preserved
under these flows: We have, infinitesimally,
\begin{eqnarray}
& & \frac{\partial}{\partial t_{m}}({\{u_{r}(z),u_{s}(z')\}}
-k_{rs}(z)\delta (z-z')) \nonumber\\
&=& {\{ {\{u_{r}(z), Q_{m}\}}, u_{s}(z') \}} + {\{u_{r}(z),
{\{u_{s}(z'), Q_{m}\}}\}} - {\{k_{rs}(z)\delta (z-z'), Q_{m}\}} \nonumber\\
&=& {\{{\{u_{r}(z),u_{s}(z')\}}, Q_{m}\}}
- {\{k_{rs}(z)\delta (z-z'), Q_{m}\}} = 0.
\end{eqnarray}
We also notice that these symmetries are abelian to each other,
in the sense that the KP flow in one charge direction is invariant
under the transformation generated by another $\hat{W}_{\infty}$
charge: i.e.,
\begin{eqnarray}
& & \delta_{m}(\frac{\partial u_{r}}{\partial t_{n}}-{\{u_{r}, Q_{n}\}})
= \epsilon_{m} {\{\frac{\partial u_{r}}{\partial t_{n}}, Q_{m}\}}
- {\{\delta_{m}u_{r}, Q_{n}\}} \nonumber\\
&=&\epsilon_{m}{\{ {\{u_{r}, Q_{n}\}}, Q_{m} \}}
- \epsilon_{m}{\{ {\{u_{r}, Q_{m}\}}, Q_{n} \}}
{}~=~ \epsilon_{m}{\{ u_{r}, {\{Q_{n}, Q_{m}\}} \}} ~=~ 0.
\end{eqnarray}
These properties are consequences of eqs.(5.2) and (5.4); especially,
eqs.(5.7) and (5.8) verify the compatibility of the KP flows. We note
that these results hold for the $W_{1+\infty}$ charges as well,
simply because of the recursion relations (2.43) or (2.40).

Furthermore, the free boson realization (3.8)-(3.9) or the bosonized
parafermion generalization (4.3)-(4.4) of the
$\hat{W}_{\infty}$ currents allows us to have a natural
representation of the Hamiltonian functions (5.1) in terms of
$\bar{j}$ and $j$. The first few currents read
\begin{eqnarray}
H_{1} &=& u_{0} ~=~ \bar{j}j, \nonumber\\
H_{2} &=& u_{1}+\frac{1}{2}u_{0}' ~=~ \frac{1}{2}(\bar{j}'j-\bar{j}j')
+\bar{j}^{2}j+\bar{j}j^{2}, \nonumber\\
H_{3} &=& u_{2}+u_{1}'+\frac{1}{3}u_{0}''+u_{0}^{2} ~=~ \frac{1}{3}
(\bar{j}''j-\bar{j}'j'+\bar{j}j'') \nonumber\\
& & \bar{j}\bar{j}'j+\bar{j}'j^{2}-\bar{j}^{2}j'-\bar{j}jj'+\bar{j}^{3}j
+3\bar{j}^{2}j^{2}+\bar{j}j^{3},
\end{eqnarray}
where $H_{1}=\hat{W}_{2}$ is the energy-momentum tensor in the classical
$SL(2,R)/U(1)$ model. Using the corresponding expressions of
the charges $Q_m$ in terms of $\bar{j}$ and $j$, one can easily summarize
the KP flows (5.5) for $u_r$ by the following flows of $j$ and $\bar{j}$:
\footnote[1]{Some recent discussions on this
issue based on eq.(3.8) are given in ref.[34].}
\begin{eqnarray}
\frac{\partial \bar{j}}{\partial t_{m}} = {\{\bar{j}, Q_{m}\}}, ~~~~
\frac{\partial j}{\partial t_{m}} = {\{j, Q_{m}\}}
\end{eqnarray}
with $Q_{m}$ here the integrals of eq.(5.9).
The first few equations in this hierarchy read
\begin{eqnarray}
& & \frac{\partial \bar{j}}{\partial t_{1}} = \bar{j}', ~~~~
\frac{\partial j}{\partial t_{1}} = j', \nonumber\\
& & \frac{\partial \bar{j}}{\partial t_{2}} = 2(\bar{j}j)'
+(\bar{j}^{2})' +\bar{j}'', \nonumber\\
& & \frac{\partial j}{\partial t_{2}} = 2(\bar{j}j)' +(j^{2})' -j''.
\end{eqnarray}
One may use the equations (5.10)
to define an integrable hierarchy whose hamiltonian structure is
simply the Poisson brackets (3.4) for $\bar{j}$ and $j$. From
the hierarchy (5.10), one can obtain all the composite KP evolutions
(5.5) for $u_{r}$. We may call this hierarchy as the two-boson
reduced KP hierarchy. (See also ref.[34].)

Similarly, the $\bar{j}$-$j$ hierarchy (5.10) is invariant under the
$\hat{W}_{\infty}$ transformations
\begin{eqnarray}
\delta_{m}\bar{j} = \epsilon_{m} {\{\bar{j}, Q_{m}\}},~~~~
\delta_{m}j = \epsilon_{m} {\{j, Q_{m}\}};
\end{eqnarray}
and the fundamental brackets (3.4) (and further the composite
$\hat{W}_{\infty}$ brackets) are invariant under the
$\bar{j}$-$j$ flows:
e.g.:
\begin{eqnarray}
\frac{\partial {\{\bar{j},j\}}}{\partial t_{m}} = {\{ {\{\bar{j},Q_{m}\}},
j\}} + {\{\bar{j}, {\{j,Q_{m}\}} \}} = {\{ {\{\bar{j},j\}}, Q_{m} \}} = 0.
\end{eqnarray}

Incidentally we point out that based on the free fermion
realization (3.2) of $W_{1+\infty}$, we have alternative
free field expressions for the KP Hamiltonians (5.1); e.g.,
\begin{eqnarray}
& & H_{1} = \bar{\psi}\psi, ~~~~ H_{2} = \frac{1}{2}(\bar{\psi}'\psi-
\bar{\psi}\psi'), \nonumber\\
& & H_{3} = \frac{1}{3}(\bar{\psi}''\psi-\bar{\psi}'\psi'+\bar{\psi}\psi'').
\end{eqnarray}
Their charges can also be used to generate a compatible and integrable
$\bar{\psi}$-$\psi$ hierarchy
\begin{eqnarray}
\frac{\partial \bar{\psi}}{\partial t_{m}} = {\{\bar{\psi}, Q_{m}\}}, ~~~~
\frac{\partial \psi}{\partial t_{m}} = {\{\psi, Q_{m}\}}.
\end{eqnarray}
It has similar invariance properties to those of the $\bar{j}$-$j$ hierarchy.
Exactly the same way, we may also have a parafermion reduced KP hierarchy
-- $\psi_{+}$-$\psi_{-}$ flows according to eq.(4.12).

To conclude, we make two remarks. First,
our results in this paper may have interesting applications to physics,
in particular to string theory. By now it is well-known that
the noncompact $SL(2,R)/U(1)$ model provides us a world-sheet sigma model
for the 2D string theory with black hole interpretation [35]. It is
conceivable that the infinite $\hat{W}_{\infty}$ symmetry or charges
would be useful in the discussion on the spectrum of 2D black holes [36].
Secondly, though our treatment in this paper is restricted to
the classical theory, it gives us some crucial hints, as we will
see in the subsequent paper [5], about how to construct a quantum
deformation of the KP hierarchy.

\vspace{30 pt}
\begin{center}
{\bf Acknowledgement}
\end{center}

We thank I. Bakas for useful discussions.
The work was supported in part by U.S. NSF-grant PHY-9008452.

\vspace{40 pt}
\begin{center}
{\large REFERENCES}
\end{center}
\begin{itemize}
\vspace{5 pt}

\item[1.] F. Yu and Y.-S. Wu, Phys. Lett. B263 (1991) 220.
\item[2.] F. Yu and Y.-S. Wu, Nucl. Phys. B373 (1992) 713.
\item[3.] F. Yu and Y.-S. Wu, Phys. Rev. Lett. 68 (1992) 2996.
\item[4.] F. Yu and Y.-S. Wu, Utah preprint UU-HEP-92/11, May 1992, to be
published in Phys. Lett. B.
\item[5.] F. Yu and Y.-S. Wu, Utah preprint UU-HEP-92/12, August 1992.
\item[6.] J. L. Gervais, Phys. Lett. B160 (1985) 277; B. A. Kuperschmidt,
Phys. Lett. A109 (1985) 417.
\item[7.] A. A. Belavin, A. M. Polyakov and A. B. Zamolodchikov, Nucl. Phys.
B241 (1984) 333; Vl. S. Dotsenko and V. A. Fateev, Nucl. Phys. B240 [FS 12]
(1984) 312; B. L. Feigin and D. B. Fuchs, Func. Aual. Appl. 16 (1982).
\item[8.] A. B. Zamolodchikov, Theor. Math. Phys. 65 (1985) 1205.
\item[9.] I. M. Gelfand and L. A. Dickey, Russ. Math. Surv. 30 (1975) 77;
Funct. Anal. Appl. 10 (1976) 259; I. M. Gelfand and I. Dorfman, Funct.
Anal. Appl. 15 (1981) 173; B. A. Kupershmidt and G. Wilson, Invent. Math.
62 (1981) 403.
\item[10.] V. A. Fateev and S. L. Lykyanov, Int. J. Mod. Phys. A3 (1988)
507; A. A. Belavin, Adv. Study Pure Math. 19 (1989) 117.
\item[11.] I. Bakas, Phys. Lett. B228 (1989) 57; Comm. Math. Phys. 134
(1990) 487.
\item[12.] C. Pope, L. Romans and X. Shen, Phys. Lett. B236 (1990) 173;
Nucl. Phys. B339 (1990) 191.
\item[13.] C. Pope, L. Romans and X. Shen, Phys. Lett. B242 (1990) 401.
\item[14.] I. Bakas and E. Kiritsis, Nucl. Phys. B343 (1990) 185.
\item[15.] M. Sato, RIMS Kokyuroku 439 (1981) 30; E. Date, M. Jimbo, M.
Kashiwara and T. Miwa, in Proc. of RIMS Symposium on Nonlinear Integrable
Systems, eds. M. Jimbo and T. Miwa, (World Scientific, Singapore, 1983);
G. Segal and G. Wilson, Publ. IHES 61 (1985) 1.
\item[16.] D. Gross and A. Migdal, Phys. Rev. Lett. 64 (1990) 127;
M. Douglas and S. Shenker, Nucl. Phys. B335 (1990) 635; E. Brezin and
V. Kazakov, Phys. Lett. B236 (1990) 144.
\item[17.] E. Witten, Nucl. Phys. B340 (1990) 281; R. Dijkgraaf and
E. Witten, Nucl. Phys. B342 (1990) 281.
\item[18.] M. Douglas, Phys. Lett. B238 (1990) 176; T. Banks, M. Douglas,
N. Seiberg and S. Shenker, Phys. Lett. B238 (1990) 279.
\item[19.] E. Verlinde and H. Verlinde, Nucl. Phys. B348 (1991) 457;
J. Goeree, Nucl. Phys. B358 (1991) 737.
\item[20.] K. Yamagishi, Phys. Lett. B259 (1991) 436.
\item[21.] Y. Watanabe, Ann. di Mat. Pura Appl. 86 (1984) 77.
\item[22.] L. A. Dickey, Annals New York Academy of Sciences, 491 (1987)
131.
\item[23.] J. M. Figueroa-O'Farrill, J. Mas and E. Ramos, Phys. Lett.
B266 (1991) 298; preprint BONN-HE-92/20, US-FT-92/7 or KUL-TF-92/20.
\item[24.] B. B. Kadomtsev and V. I. Petviashvili, Sov. Phys. Doklady
15 (1970) 539.
\item[25.] F. Yu, Nucl. Phys. B375 (1992) 173.
\item[26.] F. Yu, J. Math. Phys. 33 (1992) 3180.
\item[27.] V. Drinfel'd and V. Sokolov, Sov. Probl. Mat. 24 (1984) 81.
\item[28.] V. Knizhnik and A. B. Zamolodchikov, Nucl. Phys. B247 (1984) 83;
D. Gepner and E. Witten, Nucl. Phys. B278 (1986) 493.
\item[29.] E. Witten, Comm. Math. Phys. 92 (1986) 455.
\item[30.] P. Goddard, A. Kent and D. Olive, Phys. Lett. B152 (1985) 88;
Comm. Math. Phys. 103 (1986) 105.
\item[31.] L. Dixon, J. Lykken and M. Peskin, Nucl. Phys. B235 (1989) 215.
\item[32.] A. Gerasimov, A. Marshakov and A. Morozov, Nucl. Phys. B328
(1989) 664.
\item[33.] O. Hern\'andez, Phys. Lett. B233 (1989) 355.
\item[34.] D. A. Depireux, Laval preprint LAVAL-PHY-21-92;
J. M. Figueroa-O'Farrill, J. Mas and E. Ramos,
preprint BONN-HE-92/17, US-FT-92/4 or KUL-TF-92/26.
\item[35.] E. Witten, Phys. Rev. D44 (1991) 314.
\item[36.] J. Ellis, N. Mavromatos and D. Nanopoulos, Phys. Lett. B272
(1991) 261; B276 (1992) 56; B284 (1992) 27.

\end{itemize}

\end{document}